\documentclass[pre, reprint, letterpaper]{revtex4-1}
\usepackage{amsmath}
\usepackage{amssymb}
\usepackage{graphicx}

\begin{document}
\title{Modeling Smectic Layers in Confined Geometries: Order Parameter and Defects}
\author{Mykhailo Y. Pevnyi}
\author{Jonathan V. Selinger}
\email{jselinge@kent.edu}
\affiliation{Liquid Crystal Institute, Kent State University, Kent, Ohio 44242, USA}
\author{Timothy J. Sluckin}
\email{T.J.Sluckin@soton.ac.uk}
\affiliation{School of Mathematics, University of Southampton, Southampton SO17 1BJ, United Kingdom}
\date{July 9, 2014}

\begin{abstract}
We identify problems with the standard complex order parameter formalism for smectic-A (SmA) liquid crystals, and discuss possible alternative descriptions of smectic order.  In particular, we suggest an approach based on the real smectic density variation rather than a complex order parameter.  This approach gives reasonable numerical results for the smectic layer configuration and director field in sample geometries, and can be used to model smectic liquid crystals under nanoscale confinement for technological applications.
\end{abstract}

\maketitle

\section{Introduction}

For over forty years, theoretical understanding of smectic liquid crystals has been based on the complex order parameter $\psi(\mathbf{r})$ introduced by de Gennes~\cite{de-Gennes-SSC-1972,de-Gennes-Physics-of-LCs-1974}, which represents the magnitude and phase of layer ordering.  During this time, the order parameter has been useful in many ways.  It demonstrated an analogy between smectic liquid crystals and superconductors, allowing methods of solid-state physics to be applied to liquid-crystal science~\cite{de-Gennes-SSC-1972,Halperin-Lubensky-1974}.  It led to theories for the nematic-SmA and isotropic-SmA transitions, which are strongly affected by nematic order fluctuations~\cite{Halperin-Lubensky-Ma-1974,Chen-Lubensky-Nelson-1978,Mukherjee-EPJ-2001,Mukherjee-JCP-2004,Biscari-PRE-75-2007}.  It further led to prediction of twist-grain-boundary phases, liquid-crystal analogues of the Abrikosov flux lattice in type-II superconductors~\cite{Renn-Lubensky-1988}.  Most recently, it has led to calculations for smectic layer configurations in confined geometries~\cite{Abukhdeir-SSP-2008,Abukhdeir-NJP-2008,Abukhdeir-Macromol-2009,Abukhdeir-Langmuir-2009,Abukhdeir-LiqCryst-2009,Soule-Macromol-2009,Abukhdeir-SoftMatter-2010}, which may be useful for design of smectic devices~\cite{Spillmann}.

The purpose of this paper is to point out two problems with the complex order parameter description, which affect some but certainly not all of the work that has been done with it.  The first problem is related to the topology of the order parameter itself.  If the order parameter is treated as a single-valued complex-number field, then it is unable to describe certain types of defects that can realistically occur.  As a result, calculations based on this order parameter can predict unphysical configurations of smectic layers.  One possible solution to this problem is to regard the order parameter as a double-valued complex-number field, as has recently been proposed~\cite{Chen-PNAS-2009,Alexander-PRL-2010,Alexander-PRL-2012}.  That solution is mathematically and physically valid, but it is not well-suited to numerical calculations of smectic layer configurations.  It would be useful to find an alternative approach that could be more suitable for computation.

The second problem is related to the free energy.  The functional constructed by de Gennes represents the free energy on a coarse-grained basis, on length scales much greater than the smectic layer spacing.  It does not represent the local free energy density on the length scale of the smectic layers themselves.  As a result, it is suitable for macroscopic calculations, but not for nanoscale calculations of the positions of defects with respect to smectic layers, or the positions of smectic layers with respect to boundaries.

As a solution to these problems, we propose to use the physical density variation $\delta\rho(\mathbf{r})$ instead of the complex order parameter $\psi(\mathbf{r})$.  We develop a theory for smectic layering in terms of $\delta\rho(\mathbf{r})$, which is less mathematically elegant than the theory in terms of $\psi(\mathbf{r})$, but is suitable for numerical computation.  Through symmetry arguments and explicit calculations, we show that this theory avoids both of the problems outlined above.  As examples, we present calculations of disclination structures, including defect charges of $+1/2$, $+1$ and $\pm2$ (which have recently been studied using topological methods~\cite{Chen-PNAS-2009}).  We also present calculations of dislocation structures, showing the Peierls-Nabarro barrier for dislocation glide~\cite{Kleman-Lavrentovich}.  The results are physically reasonable, and show that the theory in terms of $\delta\rho(\mathbf{r})$ is appropriate for modeling smectic layer configurations.

\section{Problems}

\subsection{Order parameter}

\begin{figure}
\includegraphics[width=2.8in]{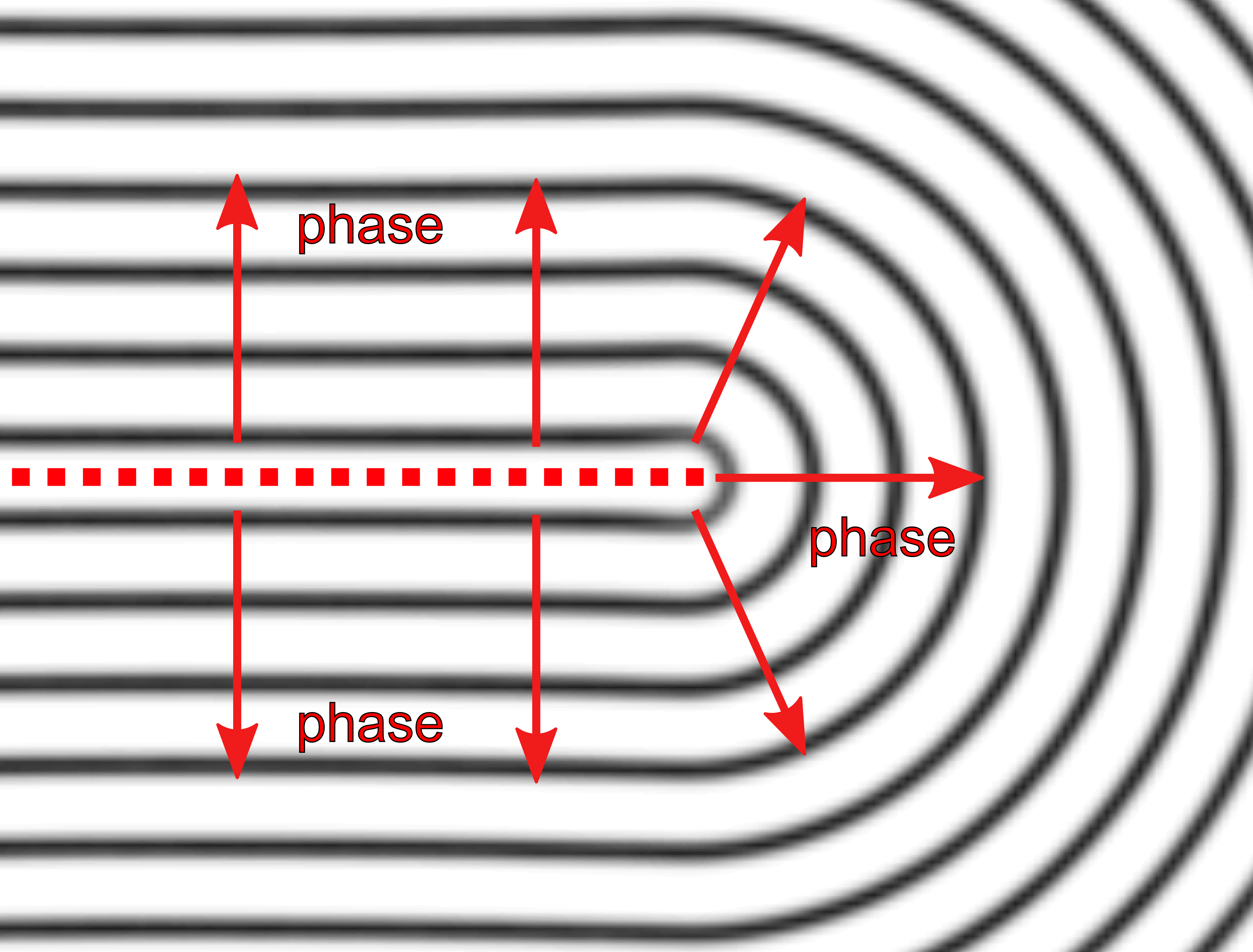}
\caption{(Color online) Disclination in a two-dimensional smectic phase. Bright and dark regions correspond to higher and lower density, respectively, and the red dotted line is the branch cut in $\psi$ and $\hat{\mathbf{n}}$.}
\label{figure-disclination-line}
\end{figure}

To see the first problem with the order parameter, consider a disclination of charge $+1/2$, as shown in Fig.~\ref{figure-disclination-line}. In this figure, every point on the plane has a local density $\rho(\mathbf{r})=\rho_0+\delta\rho(\mathbf{r})$, with bright and dark regions corresponding to higher and lower density, respectively. To use the complex order parameter $\psi(\mathbf{r})$, we must write the local density variation, compared with the average $\rho_0$, as $\delta\rho(\mathbf{r})=Re[\psi(\mathbf{r})]$. However, it is impossible to associate a unique complex number $\psi$ with each point around the defect. If we try to make this association, then we must say that the phase of $\psi$ increases downward in the lower-left quadrant, outward in the right half, upward in the upper-left quadrant, and eventually we reach an inconsistency. There must be a \emph{branch cut} where $\psi$ changes to the complex conjugate $\psi^*$, as illustrated by the dotted line. This situation is similar to the well-known problem of describing nematic order with a unit vector $\hat{\mathbf{n}}(\mathbf{r})$: going around a half-charge disclination, there must be a branch cut where $\hat{\mathbf{n}}$ changes to $-\hat{\mathbf{n}}$.

The branch cuts in $\psi$ and $\hat{\mathbf{n}}$ occur for the same physical reason: neither of these quantities gives an exactly correct description of the symmetry of the phase. The nematic phase has orientational order along the axis represented by $\pm\hat{\mathbf{n}}$, which can be described correctly by a tensor.  The vector $\hat{\mathbf{n}}$ is often adequate as an approximate description, but the sign of $\hat{\mathbf{n}}$ does not correspond to anything physical. Hence, a branch cut in the sign of $\hat{\mathbf{n}}$ is not a physical defect, and it cannot cost any free energy. Likewise, the smectic phase has higher density at some positions and lower density at other positions, and this density variation can be described correctly by the real number $\delta\rho$ or $Re(\psi)$. The full complex number $\psi$ may be mathematically convenient as an approximate description, but $Im(\psi)$ does not correspond to anything physical. Hence, a branch cut in $Im(\psi)$ is not a physical defect, and it cannot cost any free energy.

Does this issue with $\psi$ affect any calculations?  To see some specific examples, we implement the smectic formalism in a simulation.  The original free energy density proposed by de Gennes~\cite{de-Gennes-SSC-1972} is
\begin{equation}
f= \frac{1}{2}r|\psi|^2 + \frac{1}{4}u|\psi|^4 + \frac{1}{2}C |(\nabla - iq\hat{\mathbf{n}})\psi|^2 + f_N,
\label{eqn-free-energy-de-Gennes}
\end{equation}
where $q$ is the favored wavevector of smectic order and $f_N$ the nematic free energy density.  A generalized version of this free energy from Ref.~\cite{Mukherjee-EPJ-2001}, similar to Ref.~\cite{Abukhdeir-NJP-2008}, is
\begin{eqnarray}
f&=&\frac{1}{2}\alpha|\psi|^2 + \frac{1}{4}\beta|\psi|^4  + \frac{2 b_1 -e_1}{4}|\nabla_i\psi|^2 + \frac{1}{2}b_2 |\Delta\psi|^2 \nonumber\\
&&+ \frac{3}{4} e_1 n_i n_j \nabla_i\psi \nabla_j\psi^\ast + \frac{1}{2}K(\partial_i n_j)(\partial_i n_j).
\label{eqn-free-energy-complex-op-simulation}
\end{eqnarray}
For sample calculations, we numerically minimize the free energy of Eq.~(\ref{eqn-free-energy-complex-op-simulation}) using Monte Carlo simulated annealing.  
We perform the calculation on a square lattice, where each lattice site has a director $\hat{\mathbf{n}}$ and a complex order parameter $\psi$.  The required derivatives are approximated by standard finite differences.

\begin{figure}
\includegraphics[width=\linewidth]{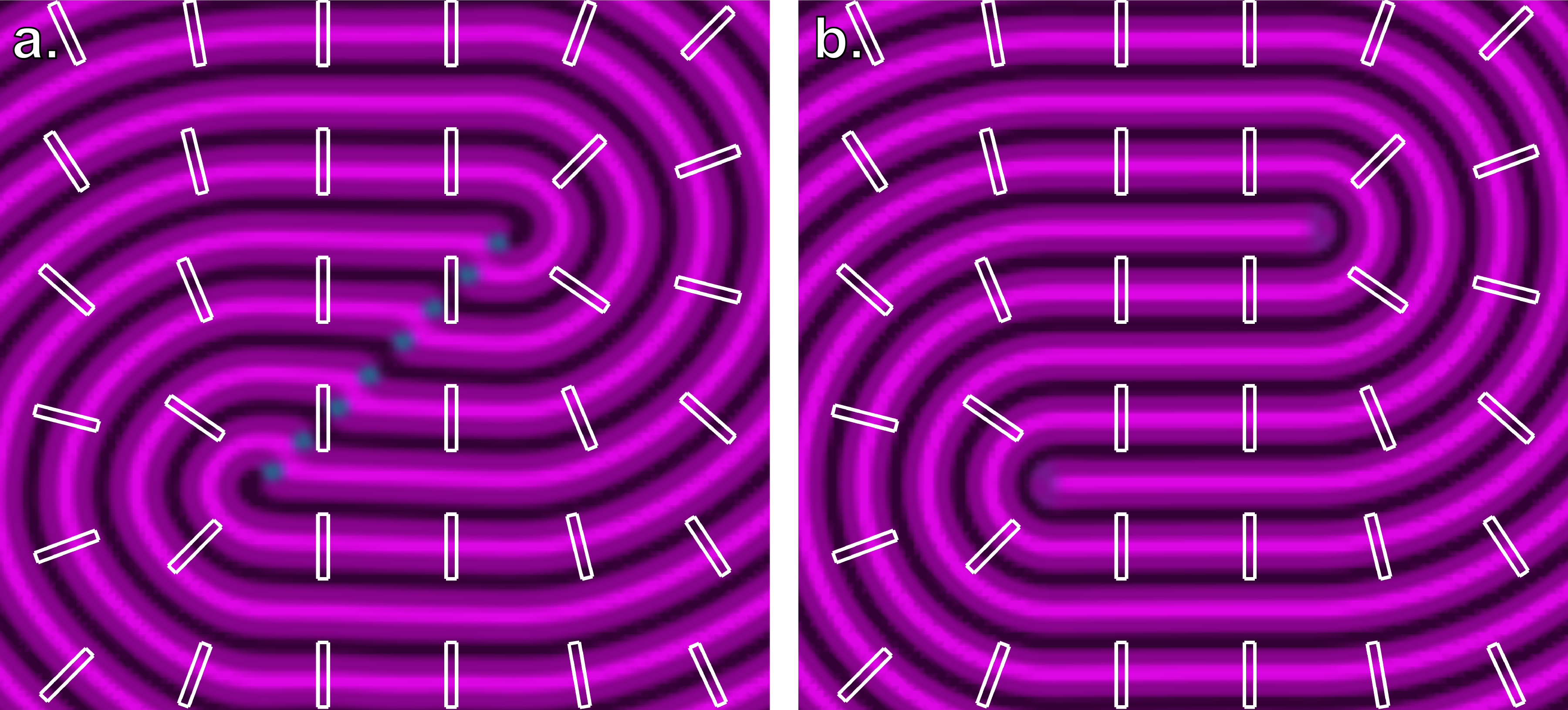}
\caption{(Color online) Simulation of smectic layers using a fixed director field (shown by short lines) with two half-charged disclinations.  (a)~Results using the complex order parameter approach of Eq.~(\ref{eqn-free-energy-complex-op-simulation}), with parameters $\alpha=-1$, $\beta=100$, $b_1=-3$, $b_2=5$, $e_1=-b_1-8 b_2 (\pi/10)^2$.  Note that the line defect between the defect cores is unphysical.  (b)~Results using the real order parameter approach of Eq.~(\ref{eqn-free-energy-real-density}), with parameters $a=-0.1$, $b=0$, $c=10$, $q=2\pi/10$, $B=0.1/q^4$.}
\label{figure-complex-vs-real-pair-disclination}
\end{figure}

For an initial simulation, we consider a geometry with two disclinations of charge $+1/2$ each.  In this initial simulation, we assume the director field is held fixed and calculate the resulting smectic layer configuration.  The results are shown in Fig.~\ref{figure-complex-vs-real-pair-disclination}(a).  Here, the color indicates the magnitude of smectic order $|\psi|$ (with purple and blue representing higher and lower order, respectively), while the brightness indicates the local density given by $Re(\psi)$.  Note that this simulation shows a line defect connecting the two disclinations.  This line defect is a sharp boundary where $\psi\to\psi^*$.  As an artifact of the model, this boundary has a free energy penalty, which is linearly proportional to the distance between disclinations and hence binds the disclinations together.

\begin{figure}
\includegraphics[width=\linewidth]{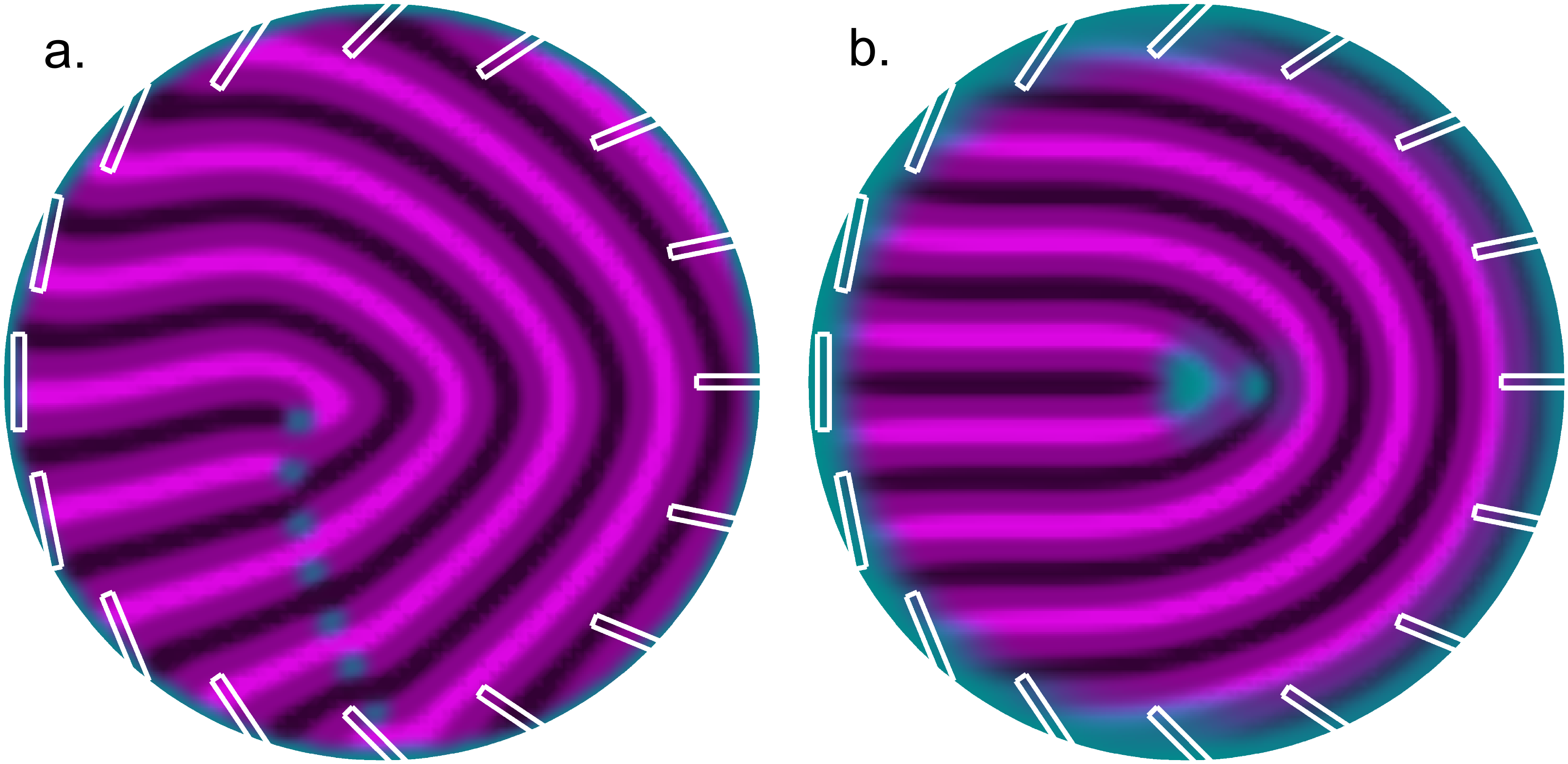}
\caption{(Color online) Simulations of a circular domain with boundary conditions requiring a single half-charged disclination.  (a)~Results using the complex order parameter approach of Eq.~(\ref{eqn-free-energy-complex-op-simulation}), with parameters $\alpha=-1$, $\beta=100$, $b_1=-3$, $b_2=5$, $e_1=-b_1-8 b_2 (\pi/10)^2$, $K= 0.0025$.  The line defect from the defect core to the boundary is unphysical.  (b)~Results using the real order parameter approach of Eq.~(\ref{eqn-free-energy-real-density}), with parameters $a=-0.1$, $b=0$, $c=10$, $q=2\pi/10$, $B=0.1/q^4$, $K=0.008$.}
\label{figure-complex-vs-real-single-disclination}
\end{figure}

For a second example, we perform simulations where the director field and layer configuration can both relax.  We consider a circular geometry with boundary conditions on the director requiring a single disclination of charge $+1/2$.  Subject to that constraint, the director and layers relax together inside the domain.  Numerical minimization of the free energy gives the structure shown in Fig.~\ref{figure-complex-vs-real-single-disclination}(a).  Once again, we see a line defect in the layers coming out of the disclination.  Because there is no other disclination where the line defect can terminate, it runs all the way to the boundary.  This line defect is not required by the symmetry of the smectic phase; it is just an artifact of the complex order parameter formalism.

One possible response to this problem is to say that complex order parameter $\psi(\mathbf{r})$ is not really a single-valued function of position.  This point may have been understood implicitly for many years.  To our knowledge, it was first stated explicitly in a recent series of papers by Kamien and collaborators~\cite{Chen-PNAS-2009,Alexander-PRL-2010,Alexander-PRL-2012}.  These papers point out that the phase of $\psi(\mathbf{r})$ is not actually an element of the unit circle $S^1$; rather, it is an element of the orbifold $S^1/\mathbb{Z}_2$.  In other words, $\psi(\mathbf{r})$ is not a single-valued function but rather a double-valued function of position.  At every position, it takes \emph{both} of the values $Re(\psi)\pm i Im(\psi)$, because these two values correspond to the same physical density.

We do not disagree with the approach of Refs.~\cite{Chen-PNAS-2009,Alexander-PRL-2010,Alexander-PRL-2012}.  Their argument is correct both mathematically and physically, and it is well-suited for some analytic calculations of smectic layer configurations.  However, that approach is not simple to implement in a numerical simulation.  For anyone who is developing software, there is a natural tendency to assume that $\psi(\mathbf{r})$ is a single complex number at each position, which is incorrect.  As a result, there is a risk that calculated structures will have unphysical line defects like the structures in Figs.~\ref{figure-complex-vs-real-pair-disclination}(a) and~\ref{figure-complex-vs-real-single-disclination}(a), and indeed such structures can be found in the literature.  To follow the approach of Refs.~\cite{Chen-PNAS-2009,Alexander-PRL-2010,Alexander-PRL-2012}, it would be necessary for software developers to construct a data structure that represents the appropriate orbifold.  As an alternative to that challenging task, it might be preferable to develop a different formalism that is not only correct but also straightforward to implement numerically.  We will propose such a formalism in Sec. III below.

\subsection{Free energy}

The second issue with the complex order parameter formalism is that it describes the free energy on \emph{coarse-grained} length scales, which are much greater than the smectic layer spacing.  For many purposes this coarse-grained description is desirable, because it allows the theory to calculate macroscopic distortions of smectic layers.  However, the coarse-grained description is not able to describe the free energy on a length scale comparable to the smectic layer spacing, and hence it cannot calculate nanoscale features of the layer configuration.

The simplest way to see the coarse-grained nature of the theory is to consider a simple periodic density wave, which is described by the complex order parameter $\psi(x,y)=e^{i q y + \Delta\Phi}$.  By putting this order parameter into the free energy density of Eq.~(\ref{eqn-free-energy-de-Gennes}) or~(\ref{eqn-free-energy-complex-op-simulation}), it is easy to see that the free energy density is constant.  All positions are equivalent, with the same free energy density, regardless of whether they are density maxima, minima, or anywhere in between.  Of coarse, the microscopic free energy density cannot really be constant; it must depend on the position with respect to the smectic layers.  The free energy density of Eq.~(\ref{eqn-free-energy-de-Gennes}) or~(\ref{eqn-free-energy-complex-op-simulation}) is just an average over the smectic density wave.

A problem occurs if one tries to use the coarse-grained free energy to calculate nanoscale properties of the smectic layers.  As a specific example, suppose we want to calculate the energy of a dislocation as a function of the position with respect to the layer structure.  This calculation would be useful to determine the most favorable position of the dislocation with respect to the layers, and to predict the Peierls-Nabarro energy barrier for dislocation glide from layer to layer (the process illustrated in Fig. 9.17 of Ref.~\cite{Kleman-Lavrentovich}).

\begin{figure}
\includegraphics[width=3.375in]{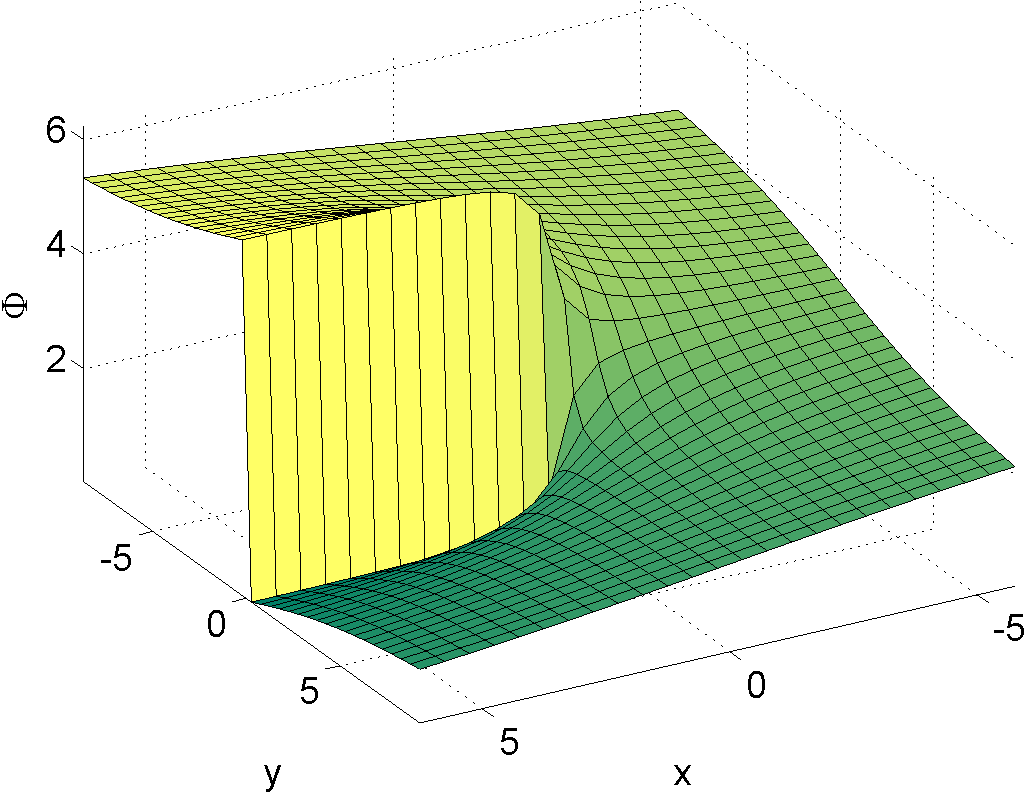}
\caption{(Color online) Visualization of the phase variation $\Phi(x,y)$ around a single edge dislocation.}
\label{dislocationphase}
\end{figure}

\begin{figure*}
\begin{tabular}{ccccc}
& $\delta\rho=Re(\psi)$ & $f(\psi)$ & $f(\rho)$: $b=0$ & $f(\rho)$: $b>0$ \\
\raisebox{0.5in}{$\Delta\Phi=0$} & \includegraphics[width=1.2in]{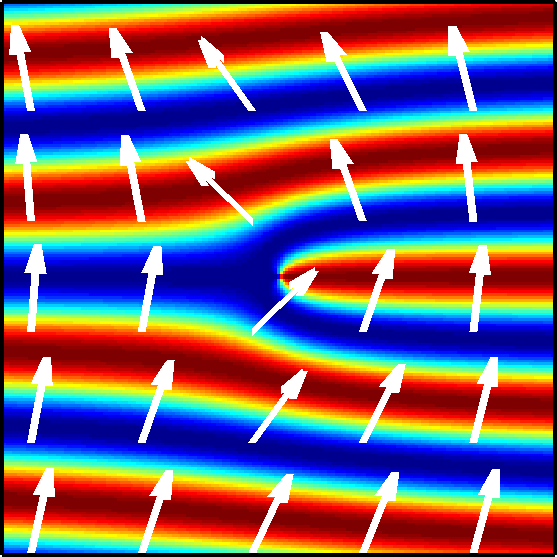} & \includegraphics[width=1.2in]{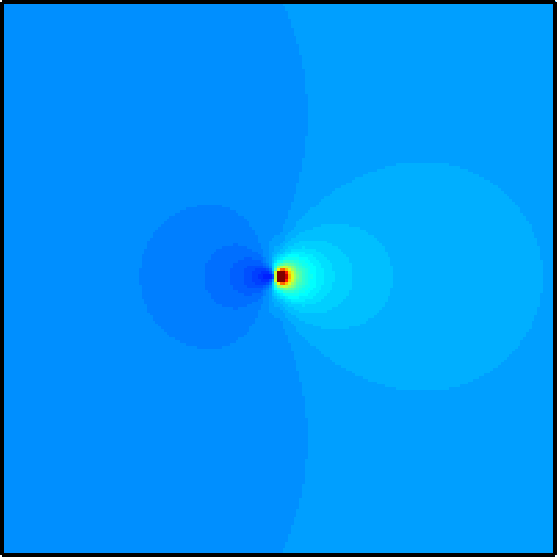} & \includegraphics[width=1.2in]{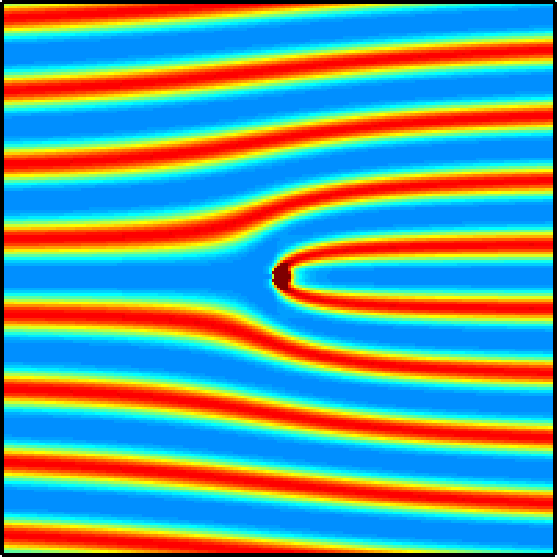} & \includegraphics[width=1.2in]{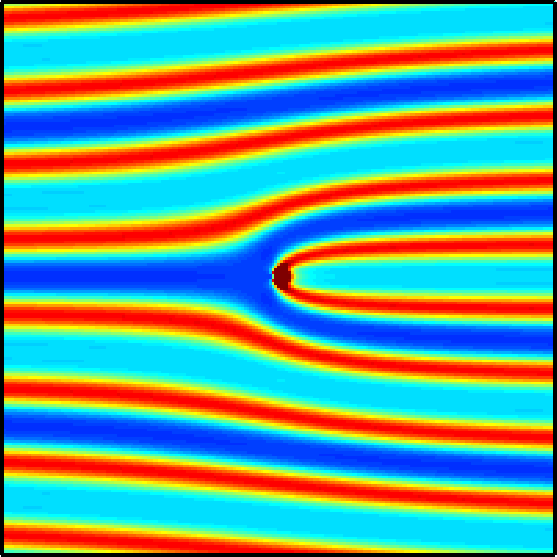} \\
\raisebox{0.5in}{$\Delta\Phi=\displaystyle\frac{\pi}{2}$} & \includegraphics[width=1.2in]{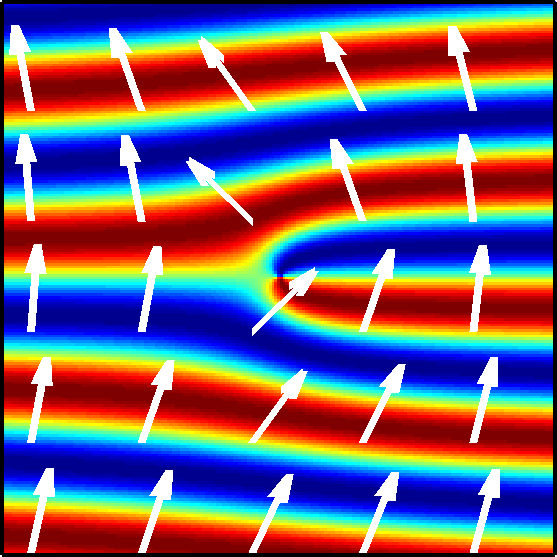} & \includegraphics[width=1.2in]{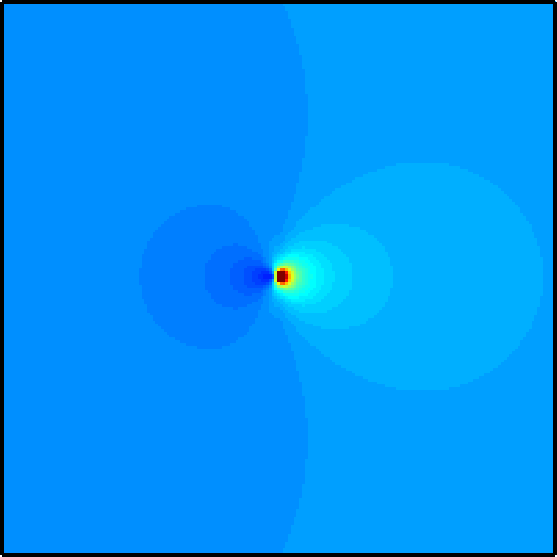} & \includegraphics[width=1.2in]{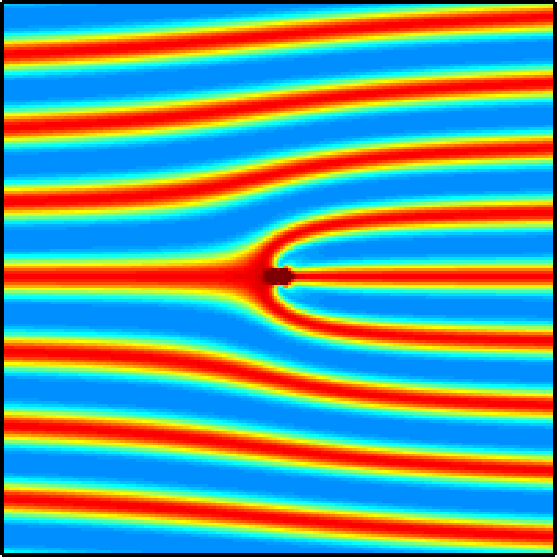} & \includegraphics[width=1.2in]{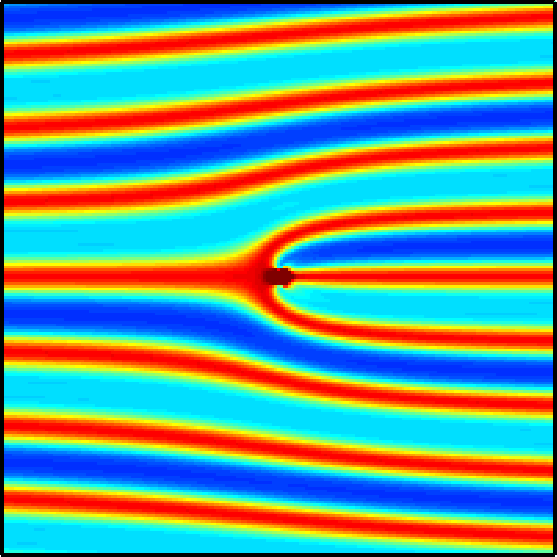} \\
\raisebox{0.5in}{$\Delta\Phi=\pi$} & \includegraphics[width=1.2in]{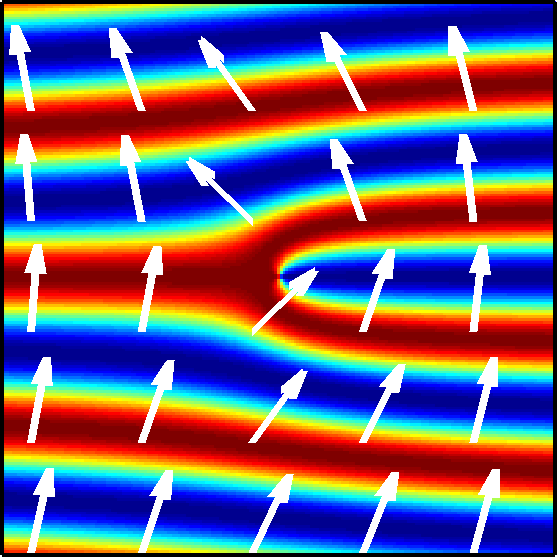} & \includegraphics[width=1.2in]{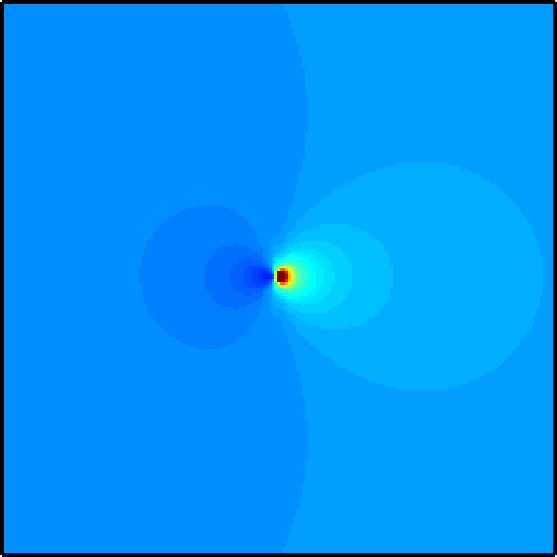} & \includegraphics[width=1.2in]{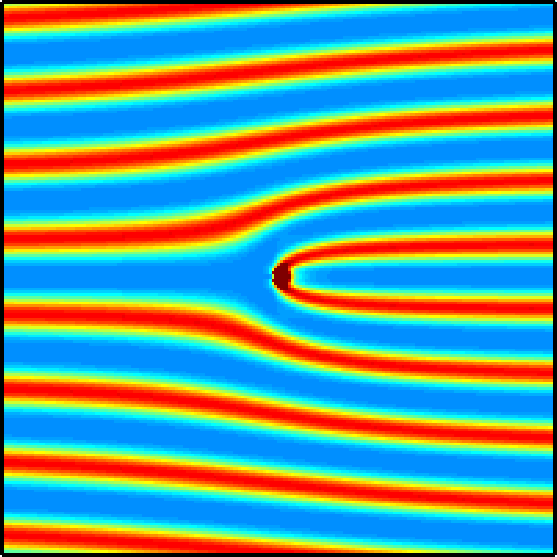} & \includegraphics[width=1.2in]{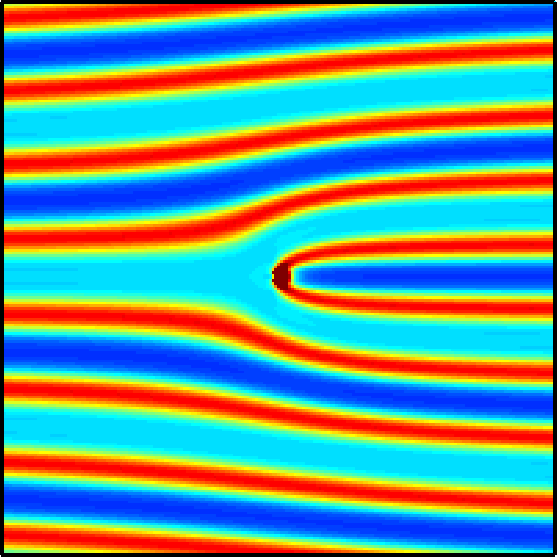} \\
\end{tabular}
\caption{(Color online) Behavior as the dislocation is displaced with respect to the layer structure.  Plots show the smectic density variation, the free energy density for the complex order parameter formalism, and the free energy density for the real density formalism (calculated for the cubic parameter $b=0$ and $b\not=0$).  Red represents highest values of density or free energy; blue represents lowest values.}
\label{dislocationexamples}
\end{figure*}

To describe a single edge dislocation in the 2D $(x,y)$ plane, we use the complex order parameter $\psi(x,y)=e^{i(qy+\Phi(x,y)+\Delta\Phi)}$,
where $\Phi(x,y) = \arg(x+iy)$ is the phase variation and $\Delta\Phi$ is a constant phase offset.  The director $\mathbf{\hat{n}}(x,y)$ is chosen as a unit vector along the gradient of $\psi(x,y)$.  Figure~\ref{dislocationphase} shows a visualization of $\Phi(x,y)$; the branch cut starts at $(x=0, \, y=0)$ and goes in the positive $x$ direction.

The first column of pictures in Fig.~\ref{dislocationexamples} shows visualizations of the density variation around the dislocation.  Note that the constant phase offset $\Delta\Phi$ defines the position of the dislocation with respect to the layer structure.  When $\Delta\Phi=0$ the dislocation occurs at a density minimum (shown in blue); when $\Delta\Phi=\pi$ it occurs at a density maximum (shown in red).  For intermediate $\Delta\Phi$ it occurs at a lower-symmetry point between those extremes.

The second column of pictures in Fig.~\ref{dislocationexamples} shows the free energy density of Eq.~(\ref{eqn-free-energy-de-Gennes}), calculated with the parameters $r=-5$, $u=5$, $q=1$, and $C=1/q^2$.  From these pictures, we can see that the free energy density is sharply peaked at the dislocation, and decays rapidly away from the dislocation.  This free energy density is clearly independent of the constant phase offset $\Delta\Phi$, and indeed all the pictures in this column are identical.  As a result, the integrated free energy is also independent of $\Delta\Phi$, and hence the dislocation is equally likely to occur anywhere within the layer structure.  This result implies that the dislocation can move with respect to the layer structure, from row to row in the table, with no energy cost.  In the terminology of dislocation theory, we would say that the Peierls-Nabarro energy barrier for dislocation glide is zero, which is physically unrealistic.

The same type of problem could occur in any calculations where the phase of the smectic layer structure is important, such as a calculation of the positions of smectic layers with respect to boundaries.  It shows that the model has more symmetry than the actual SmA phase:  in the real system, the density maxima and minima are special positions, and there must be some free energy difference between defects/boundaries at those positions and at other positions.  Hence, in order to calculate nanoscale features of the layer configuration, it is necessary to develop a different theoretical formalism.

\section{Proposed solution}

In the previous section, we pointed out problems in using the complex order parameter approach for nanoscale calculations of smectic layer configurations.  In this section, we consider possible solutions.

For a first possible solution, we might want to make a minimal modification of the complex order parameter approach to avoid the problem of a double-valued order parameter.  For this modification, we can replace the double-valued complex order parameter
\begin{equation}
\psi(\mathbf{r})=|\psi|e^{\pm i\phi(\mathbf{r})}=|\psi|(\cos\phi(\mathbf{r})\pm i\sin\phi(\mathbf{r}))
\end{equation}
by the single-valued complex order parameter
\begin{eqnarray}
\tilde{\psi}(\mathbf{r})&=&|\psi|e^{i|\phi(\mathbf{r})|}=|\psi|(\cos\phi(\mathbf{r})+i|\sin\phi(\mathbf{r})|)\nonumber\\
&=&|\psi|\left[\cos\phi(\mathbf{r})+i\sqrt{1-\cos^2 \phi(\mathbf{r})}\right].
\end{eqnarray}
The local density is related to the order parameter by
\begin{eqnarray}
\delta\rho(\mathbf{r})&=&\rho(\mathbf{r})-\rho_0=Re[\psi(\mathbf{r})]=Re[\tilde{\psi}(\mathbf{r})]\nonumber\\
&=&|\psi|\cos\phi(\mathbf{r}),
\end{eqnarray}
and the amplitude of the density modulation is
\begin{equation}
\delta\rho_\text{max}=|\psi|.
\end{equation}
The single-valued complex order parameter can then be written as
\begin{equation}
\tilde{\psi}(\mathbf{r})=\delta\rho(\mathbf{r})+i\sqrt{\delta\rho_\text{max}^2 - \delta\rho(\mathbf{r})^2}.
\end{equation}
In that case, the free energy density of Eq.~(\ref{eqn-free-energy-de-Gennes}) can be transformed into
\begin{eqnarray}
f&=&\frac{r}{2}|\tilde{\psi}|^2 + \frac{u}{4}|\tilde{\psi}|^4 + \frac{C}{2} \left|(\nabla - iq\hat{\mathbf{n}})\tilde{\psi}\right|^2 + f_N \nonumber\\
\label{free-energy-complex-single-valued}
&=&\frac{r}{2}\delta\rho^2 + \frac{u}{4}\delta\rho^4\\
&&+\frac{C \delta\rho^2 \left|\nabla(\delta\rho)+q\hat{\mathbf{n}}\sqrt{\delta\rho_\text{max}^2 - \delta\rho^2}\right|^2}%
{2(\delta\rho_\text{max}^2 - \delta\rho^2)} + f_N . \nonumber
\end{eqnarray}

In principle, it should be possible to use the free energy density of Eq.~(\ref{free-energy-complex-single-valued}) to calculate smectic layer configurations, without any problem with a double-valued order parameter.  However, we find this solution to be unsatisfactory for three reasons.  First, this free energy density requires knowledge of the modulation amplitude $\delta\rho_\text{max}$ as well as the local density $\delta\rho(\mathbf{r})$; that information is not always available.  Second, the denominator of Eq.~(\ref{free-energy-complex-single-valued}) will certainly cause numerical singularities wherever it is near zero.  Third, this equation does not give a free energy density that varies periodically with the smectic density wave, and hence it does not solve the problem described in Sec.~II(B).  All three of these issues arise for the same reason:  The single-valued complex order parameter $\tilde{\psi}(\mathbf{r})$ has an imaginary part that is only present for consistency with the original de Gennes formalism; it is not physically necessary or meaningful.

Because of these issues, we suggest an alternative formalism based only on the real density variation $\delta\rho(\mathbf{r})$, with no imaginary part.  In that case, the theory becomes a form of density functional theory, analogous to early work on smectic phases~\cite{Poniewierski-PRA-1991,Linhananta-PRA-1991}.  For this theory, we propose a free energy density for the SmA phase of the form
\begin{eqnarray}
f&=&\frac{a}{2}\delta\rho^2 + \frac{b}{3}\delta\rho^3 + \frac{c}{4}\delta\rho^4
+ B\left[(\partial_i \partial_j+q^2 n_i n_j )\delta\rho \right]^2 \nonumber\\
&&+ \frac{1}{2}K (\partial_i n_j)^2 ,
\label{eqn-free-energy-real-density}
\end{eqnarray}
where $\delta\rho(\mathbf{r})$ is the local deviation from the average density $\rho_0$.  This model has a transition from the nematic phase when $a$ is above a threshold to the SmA phase when $a$ is below the threshold.  In the SmA phase, the free energy minimum is approximately a sinusoidal density wave with wavelength $2\pi/q$.  This result is consistent with the de Gennes theory.

We should make four remarks about this free energy.  First, $\hat{\mathbf{n}}$ only enters through the second-rank tensor $n_i n_j$, which corresponds to $Q_{ij}$ in Refs.~\cite{Mukherjee-EPJ-2001,Mukherjee-JCP-2004,Biscari-PRE-75-2007}.  Hence, the free energy depends only on $n_i n_j$ and $\delta\rho$, which are both physical, single-valued functions, with no need for branch cuts.  Second, the free energy includes a third-order term of $\delta\rho^3$.  This term is allowed because there is no symmetry between high density ($\delta\rho>0$) and low density ($\delta\rho<0$).  Third, it could include other terms permitted by symmetry, such as $|\nabla\delta\rho|^2$ and $(\hat{\mathbf{n}}\cdot\nabla\delta\rho)^2$.  These terms shift the nematic-SmA transition and the wavelength of the smectic density modulation, but do not change the general physics discussed here, so we will not consider them further.  Fourth, it includes the nematic free energy density $f_N$.  We use the simplest approximation $f_N=\frac{1}{2}K (\partial_i n_j)^2$ with a single Frank elastic constant, although it could be generalized to different Frank constants.

\begin{figure}
\includegraphics[width=\linewidth]{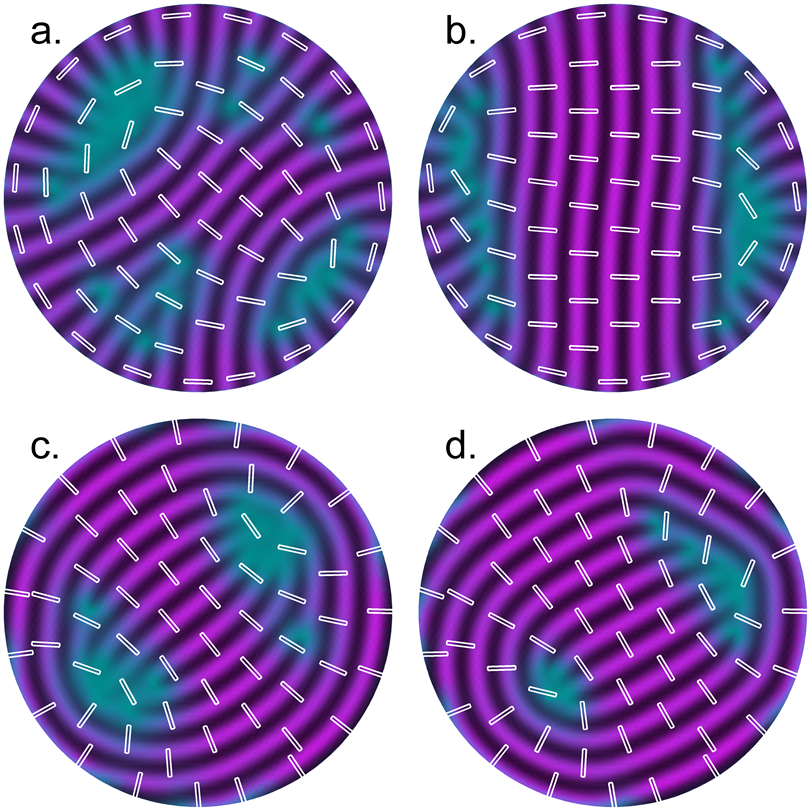}
\caption{(Color online) Simulations of a circular domain with boundary conditions requiring tangential alignment (parts a and b) or radial alignment (c and d) of the director.  The density functional free energy~(\ref{eqn-free-energy-real-density}) is used.  In a and c, parameters are $a=-5$, $b=0$, $c=5$, $B=10^{-5}$, $q=40$, $K=0.3$.  In b and d, the Frank constant is reduced to $K=0.05$.}
\label{figure-sma-circles}
\end{figure}

\begin{figure*}
\includegraphics[width=5.0in]{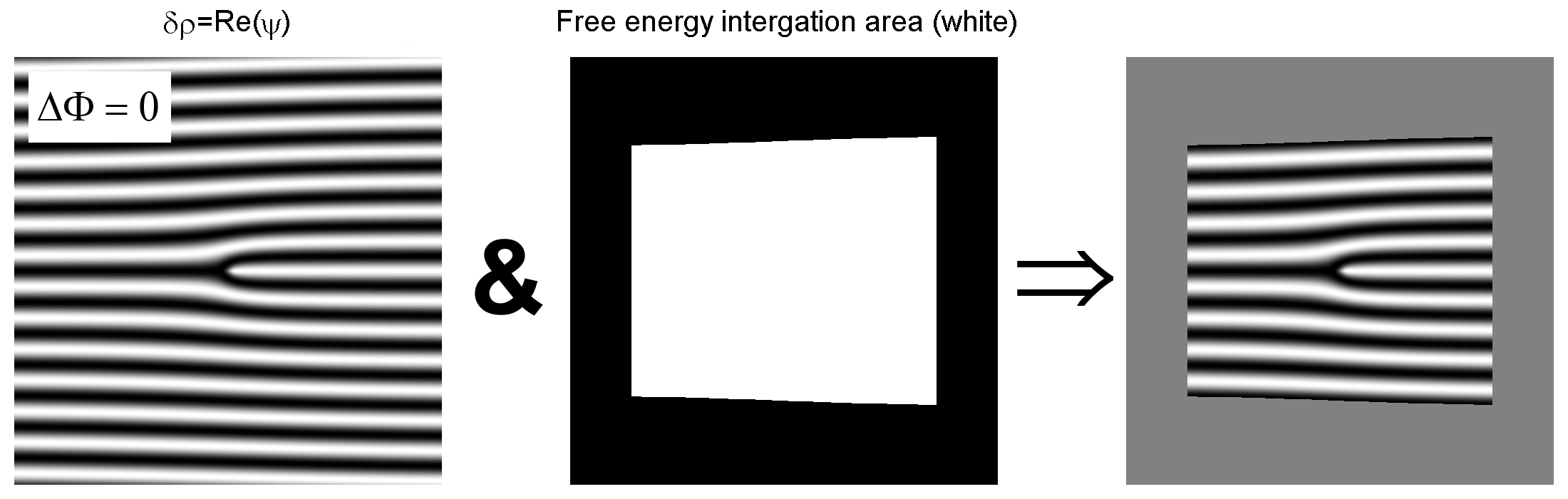}
\caption{Integration of the free energy density over a region containing an integer number of smectic layers.}
\label{integrationregion}
\end{figure*}

\begin{figure*}
\includegraphics[width=2.5in]{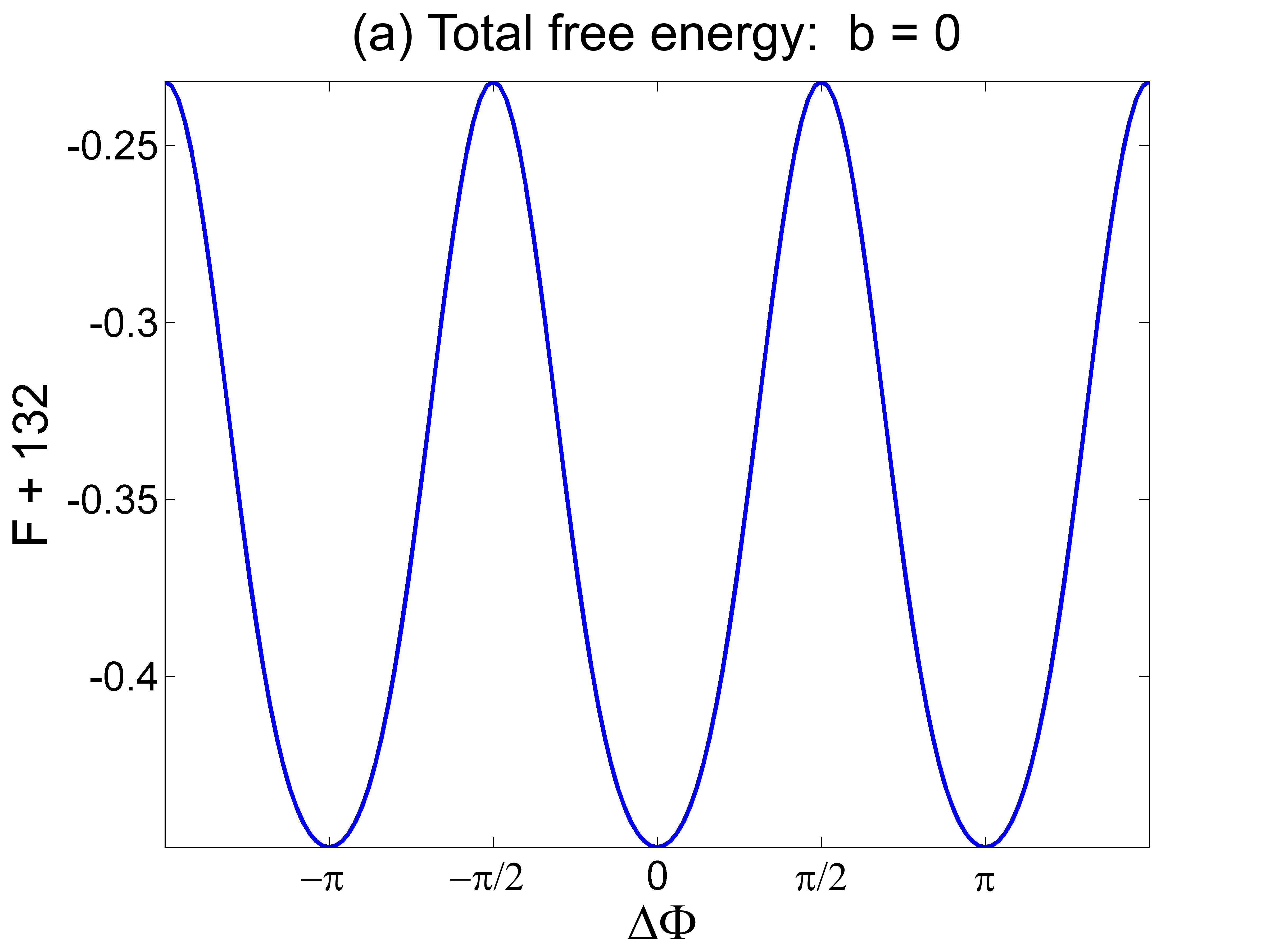}
\includegraphics[width=2.5in]{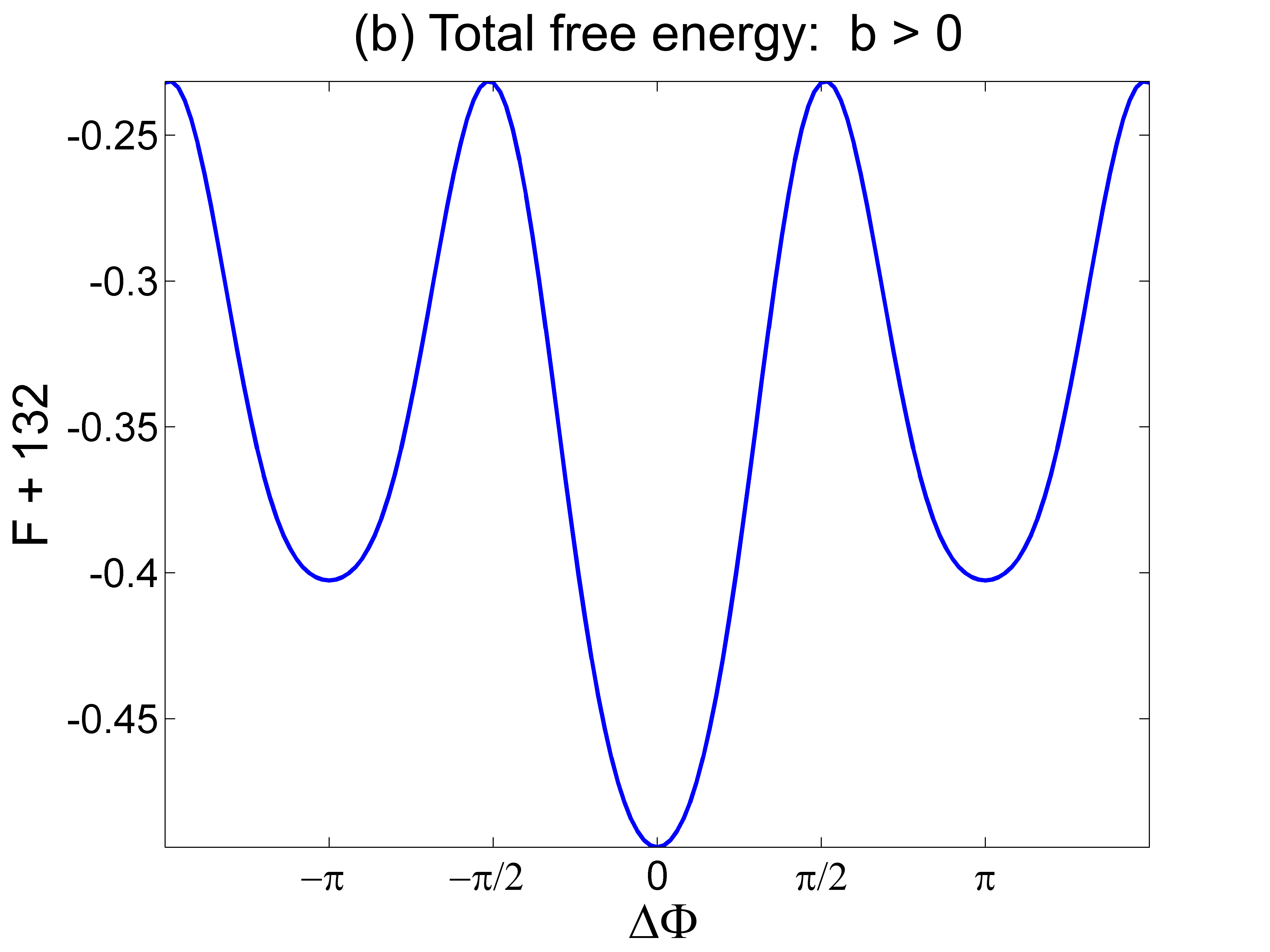}
\caption{(Color online) Integrated free energy in the real density formalism as the dislocation is displaced with respect to the layer structure, as given by the parameter $\Delta\Phi$.  (a) Parameters $a=-10$, $c=10$, $q=1$, $B=0.1/q^4$, and $b=0$, so that there is a symmetry between density minima and maxima. (b) Parameter $b=1$, breaking the symmetry between density minima and maxima.}
\label{FvsDeltaPhi}
\end{figure*}

To illustrate the physical significance of this formalism, we will consider several examples of disclinations and dislocations.  In these numerical examples, we want to describe the degree of smectic order as a function of position.  For this purpose, we need the magnitude of density modulation at the wavelength corresponding to smectic layers.  The simplest representation for this order parameter is a local Fourier transform of the density near any point, at the wave vector $q$ along the local director,
\begin{equation}
S_A(\mathbf{r})=\frac{q}{2 \pi}\left|\int_{-\pi/q}^{\pi/q} e^{-i q l} \rho(\mathbf{r}+ \mathbf{n} l) dl \right| .
\label{eqn-order-parameter}
\end{equation}
This quantity is calculated by numerical integration and presented as the smectic order parameter in the figures discussed below.

For initial tests of the real density formalism of Eq.~(\ref{eqn-free-energy-real-density}), we consider the same geometries with disclinations that were studied using the complex order parameter formalism in Sec.~II(A).  First, we consider a geometry with two disclinations of charge $+1/2$ each, assume that the director field is held fixed, and calculate the resulting smectic layer configuration.  The results are shown in Fig.~\ref{figure-complex-vs-real-pair-disclination}(b).  This structure is consistent with all the symmetries of the SmA phase.  The layers are equally spaced and normal to the director everywhere.  The region between the disclinations is a well-ordered smectic phase with no line defect.

Second, we perform simulations where both the director field and layer configuration can relax, using a circular geometry with boundary conditions on the director requiring a disclination of charge $+1/2$.  We obtain the structure shown in Fig.~\ref{figure-complex-vs-real-single-disclination}(b), which has a single disclination at the center.  The smectic layers form a relaxed configuration about the disclination.  There is a point defect in the layers at the disclination core, where we can see a reduction in the smectic order parameter defined by the local Fourier transform.  Everywhere else, the layers are well-ordered and equally spaced.

For further examples of the density functional theory, we perform simulations of the circular domains shown in Fig.~\ref{figure-sma-circles}.  Here, the director field has tangential boundary conditions (parts a and b) or radial boundary conditions (c and d).  In either case, it must have total topological charge of +1. The density modulation has free boundary conditions. We use two values of the Frank elastic constant $K$ compared to the nematic-smectic coupling $B$, and hence two values of the length scale $\lambda=(K/B)^{1/2}$.  This characteristic smectic length scale is large in a and c, and smaller in b and d.  In all cases, free energy minimization gives a configuration with two disclinations of topological charge +1/2 each, \emph{not} a single disclination of +1.  In the two cases with high $\lambda$, the director has a smooth variation between the disclinations, and the layers adapt to the director, with small variations in the layer spacing.  In the two cases with smaller $\lambda$, the layers are equally spaced over most of the domain, and the director adapts to the layers, with director variation concentrated in small regions near the boundary.  These results are physically reasonable, and correspond to what might be observed for smectic liquid crystals under nanoscale confinement~\cite{Spillmann}.

We next consider the structure and energy of a dislocation as it glides between smectic layers.  As shown in Sec.~II(B), the complex order parameter formalism describes the free energy only on a coarse-grained basis, and does not show how the energy depends on the position of a dislocation with respect to the layers.  We now repeat that dislocation calculation using the real density formalism of Eq.~(\ref{eqn-free-energy-real-density}).  The results are shown in Fig.~\ref{dislocationexamples}, next to the corresponding figures for the complex order parameter calculation.  In this figure, the third and fourth columns of pictures show the free energy density calculated for the parameters $a=-10$, $c=10$, $q=1$, $B=0.1/q^4$, and $b=0$ (in the third column) and $b=1$ (in the fourth column).  The $b$ term is important because it is the only term considered here that is odd in $\delta\rho$, and hence the only term that distinguishes between density minima and maxima.

From these images, we can make several observations.  First, the free energy density is not uniform but periodic in the smectic layer structure.  If $b=0$, there are equal free energy valleys at the density minima and maxima.  If $b>0$, the symmetry between minima and maxima is broken (as is physically realistic), and the deepest free energy valleys are at the density minima.  Furthermore, there is additional free energy associated with the dislocation itself.  Most importantly, the free energy plots change as the constant phase offset $\Delta\Phi$ is varied, i.~e.\ as the dislocation moves with respect to the layer structure.  Hence, this model does not have the unphysical symmetry found in the complex order parameter formalism.

To calculate the barrier for dislocation motion, we must integrate the free energy density to find the total free energy as a function of $\Delta\Phi$.  In this calculation, it is important to integrate over an integer number of layers, so that the result is not influenced by the number of fractional layers within the integration region.  Hence, we define the integration region by $|qx| < 10\pi$ and $|qy+\Phi(x,y)| < 8\pi$, as shown in Fig.~\ref{integrationregion}.

Figure~\ref{FvsDeltaPhi} presents graphs of the total free energy as a function of $\Delta\Phi$.  We can see that it has a periodic series of peaks and valleys as the dislocation moves with respect to the layer structure.  If $b=0$, the valleys occur whenever $\Delta\Phi$ is a multiple of $\pi$, i.~e. whenever the dislocation is at either a density minimum or maximum.  If $b>0$, the deepest valleys occur when $\Delta\Phi$ is a multiple of $2\pi$, i.~e. when the dislocation is at a density minimum; the valleys at density maxima are less deep.  The Peierls-Nabarro energy barrier for dislocation glide is then the difference in free energy between the deepest valleys and highest peaks in this plot.  Clearly this barrier is nonzero, as is physically reasonable.  We expect that this formalism would also show how the energy varies as a function of the position of layers with respect to boundaries.

As a final example, we can use the real density formalism to calculate smectic layer configurations around disclinations of total topological charge $\pm2$.  This problem is a subject of current research interest, because Kamien and collaborators~\cite{Chen-PNAS-2009} have recently used topological results of Po\'enaru to show that a smectic phase \emph{cannot} have disclinations of positive charge higher than $+1$.  By contrast, a smectic phase \emph{can} have disclinations of any arbitrarily high negative charge (integer or half-integer).  This mathematical result leads to a physical question:  If a smectic phase were put into a domain where it is forced to have a total topological charge of $+2$, how would it respond?  How could the smectic layers adapt without violating the mathematical constraint?

\begin{figure}
\includegraphics[width=\linewidth]{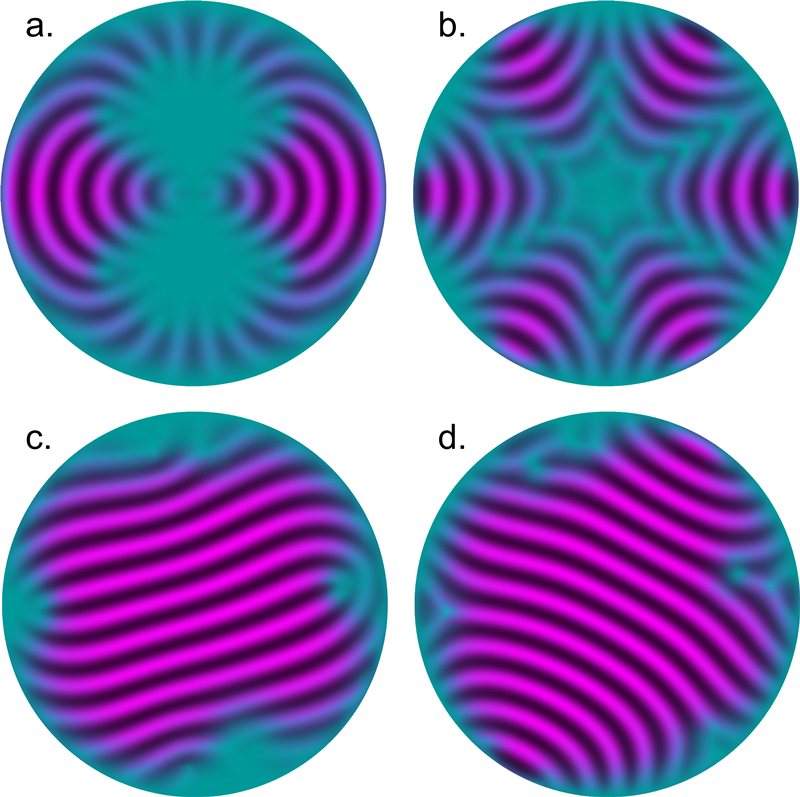}
\caption{(Color online) Simulations of smectic layer configurations around disclinations of total topological charge $+2$ (parts a and c) and $-2$ (parts b and d).  In a and b, the nematic director field inside the domain is specified, and the smectic layers respond to it.  In c and d, the boundary conditions require a total topological charge of $\pm2$, and the director field and the smectic layers inside the domain are both free to relax.}
\label{plusminus2defects}
\end{figure}

To answer that question, we simulate smectic layers in circular domains with total topological charge of $\pm2$, as shown in Fig.~\ref{plusminus2defects}.  In parts a and b, we assume that the nematic director field is fixed throughout the interior of each domain, so that there is a nematic disclination of $\pm2$ at the center.  In response to this highly charged nematic disclination, the smectic layers select a complex configuration.  For the $+2$ defect, the smectic configuration has a distribution of dislocations in the layers.  This distribution of dislocations is the response to the frustration caused by a director field that is incompatible with smectic order:  Wherever the smectic phase cannot follow the director field, it melts into dislocations.  By comparison, for the $-2$ defect, the smectic layers are highly curved but do not have the same population of dislocations.  Thus, these figures provide specific illustrations of the topological results.

In parts c and d, we impose boundary conditions on the nematic director field that require a topological charge of $\pm2$, but we do not constrain the director field inside the domain.  Instead, the director field and the smectic layers can respond together to the boundary conditions.  In each of these cases, the high topological charge breaks up into four disclinations of charge $\pm1/2$ each.  These four disclinations repel each other, and hence move near the boundaries.  Both the positive and negative disclinations are accompanied by a distribution of dislocations.  These dislocations are presumably required by topology in the positive case, and only favored by energy in the negative case, but in practice they appear fairly similar.

\section{Conclusions}

In conclusion, we have identified two problems with using the complex order parameter formalism to model nanoscale layer configurations in smectic liquid crystals.  The first problem is related to the complex order parameter itself:  If this order parameter is interpreted as a single-valued complex number field, then it is unable to describe half-charged disclinations without unphysical line defects.  This problem can be solved by reinterpreting the order parameter as a double-valued function of position, but this procedure is not well-suited for numerical simulation.  The second problem is related to the free energy:  Because the formalism uses a coarse-grained free energy that averages over the smectic layers, it does not show how the free energy depends on the position of dislocations or boundaries with respect to the layers.  This problem cannot be solved by reinterpreting the order parameter as a double-valued function.  In response to these problems, we propose an alternative formalism based on the physical density variation instead of the complex order parameter.  Through explicit calculations, we demonstrate that it gives physically reasonable results for sample geometries.  Thus, it has potential for future design of nanoscale smectic devices.

\section{Acknowledgments}

For helpful discussions we thank M. Linehan, who reached related conclusions by a different route, B.~R. Ratna, who provided experimental motivation for this study, and R. D. Kamien, who provided insight on $\pm2$ disclinations.  Some of this work was carried out at the Isaac Newton Institute for Mathematical Sciences, Cambridge, UK, which we thank for its hospitality.  This work was supported by NSF Grants DMR-0605889 and 1106014.

\bibliography{SmecticLayers-bibliography}

\end{document}